\documentclass[final,3p,times,twocolumn]{elsarticle}
\usepackage{graphicx}
\usepackage{epsfig}

\usepackage{amssymb}
\usepackage{hyperref}
\hypersetup{colorlinks=blue}
\usepackage{color}
\journal{arXiv}

\begin{document}

\begin{frontmatter}



\title{Bosonization of Cooper pairs and novel Bose-liquid superconductivity and superfluidity in high-$T_c$ cuprates
and other systems}

\author{S. Dzhumanov}

\ead{dzhumanov47@gmail.com}

\address{Institute of Nuclear Physics, Uzbek Academy of Sciences, Ulugbek, 100214, Tashkent, Uzbekistan}

\begin{abstract}
The universal criteria for bosonization of Cooper pairs and novel
Bose-liquid superconductivity and superfluidity in pseudogap
matters (high-$T_c$ cuprates and other systems with low Fermi
energies) are formulated by using the uncertainty principle and
the composite-boson mean field theory. We have established that
the often discussed $s$- or $d$-wave superconductivity occurring
in the fermionic limit of Cooper pairs can exist in conventional
superconductors (e.g., ordinary metals and heavily overdoped
cuprates) with large Fermi energies but the Fermi-liquid
(BCS-type) superconductivity is not characteristic of underdoped
to overdoped cuprates with low Fermi energies. The unusual
superconducting order parameter in high-$T_c$ cuprates and other
pseudogap matters cannot be determined as the BCS-like ($s$- or
$d$-wave) gap. We argue that many experimental data (including
tunneling and angle-resolved photoemission data) are not accurate
to identify the true superconducting order parameter in high-$T_c$
cuprates. We show that the unconventional
superconductivity/superfluidity occurring in the bosonic limit of
Cooper pairs would exist in low Fermi energy systems where the
bosonic Cooper pairs are formed at a pseudogap temperature $T^*$
above the superconducting/superfluid transition temperature $T_c$
and then part of such Cooper pairs condenses into a Bose
superfluid at $T_c$. Diamagnetism of bosonic Cooper pairs exists
in high-$T_c$ cuprates below $T^*$. Upon lowering the temperature,
the pair condensation of attracting bosons occurs first at $T_c$
and then their single particle condensation sets in at $T_c^*$
lower than $T_c$ (in three dimensions (3D)) or at $T=0$ (in two
dimensions (2D)). The coherent single particle and pair
condensates of bosons exist as two distinct superfluid phases and
arise from an effective attraction between bosons in some domains
of momentum space. By solving the mean field equations for
high-$T_c$ cuprates, the novel superconducting states (i.e., a
vortex-like state existing below the temperature $T_v=T^{2D}_c$
lower than $T^*$ but higher than $T_c=T^{3D}_c$ as well as two
superconducting phases below $T_c$) and their properties
characterized by the boson superfluid stiffness are
self-consistently determined and compared with the key
experimental findings. The mechanisms responsible for the novel
Bose-liquid superconductivity and superfluidity could be common to
a wider class of exotic superconductors/superfluids, including
quantum liquids, atomic Fermi gases and low-density nuclear
matter.
\end{abstract}

%\begin{keyword}
%%% keywords here, in the form: keyword \sep keyword
%High-$T_c$ cuprates and related systems \sep Bosonic Cooper pairs
%\sep Superconductivity and superfluidity \sep Pseudogap \sep
%Diamagnetism \sep Vortex-like state

% MSC codes here, in the form: \MSC code \sep code
% or \MSC[2008] code \sep code (2000 is the default)
%\end{keyword}
\end{frontmatter}
\section{Introduction}

The conventional superconductivity in simple metals with large
Fermi energies $\varepsilon_F>>1$ eV and small phonon energies
(Debye energies) $\hbar\omega_D$ is well described in terms of the
Bardeen-Cooper-Schrieffer (BCS) condensation of weakly-bound
(large) Cooper pairs \cite{1}. However, in more complex systems
which are of significant current interest in condensed matter
physics and beyond, our understanding of the phenomena of
superconductivity and superfluidity is still far from
satisfactory. The BCS picture, in which electrons form Cooper
pairs as a result of conventional electron-phonon interactions, is
now believed to account well for the great majority of metallic
superconductors. But there is a growing number of exotic systems,
including the unconventional high-$T_c$ cuprate superconductors
and other related systems (e.g., heavy-fermion and organic
compounds, liquid $^3$He, ultracold atomic Fermi gases and
low-density nuclear matter), in which
superconductivity/superfluidity appears anomalous and where the
origin of this phenomenon remains controversial.

After the discovery of high-$T_c$ superconductivity in doped
copper oxides (cuprates), various mechanisms have been proposed
for unconventional superconductivity, especially in the cuprates.
Many of the proposed mechanisms for unconventional cuprate
superconductivity are based on the BCS-like ($s$- or $d$- wave)
pairing correlations and on the usual Bose-Einstein condensation
(BEC) of an ideal Bose-gas of tightly-bound Cooper pairs and other
bosonic quasiparticles (e.g., bipolarons and holons)
\cite{2,3,4,5,6,7,8,9,10,11,12,13}. In these two limiting cases
the appearance of superconductivity in the BCS-like and BEC
regimes is assumed possible. However, a very controversial
question in the theory of superconductivity is the relation of
both the BCS-like pairing of carriers and the usual BEC of small
Cooper pairs to the unconventional superconductivity. The BCS-type
Cooper pairing may have a certain relation to the unconventional
superconductivity, so it is necessary to study the formation of
Cooper pairs with the determination of their genuine nature, which
is very important in the establishment of the following key
scenarios of superconducting phase transitions. In some versions
of the theory of high-$T_c$ cuprate superconductivity
\cite{2,3,9}, the crossover from BCS-like pairing regime to
real-space pairing or BEC regime \cite{14,15,16} is considered.
Renewed interest in this crossover arose with the study of the
anomalous behavior of high-$T_c$ materials. Unlike conventional
weak-coupling BCS superconductors, the high-$T_c$ cuprate
compounds falling between the BCS and BEC limits exhibit the new
physics and are characterized by small Cooper pairs, which may
have the bosonic nature. For these reasons, the BCS-Eliashberg
theory, which is a very good approximation for ordinary metals,
turned out to be inadequate for the description of unconventional
superconductivity in high-$T_c$ cuprates where the Fermi energy
$\varepsilon_F$ becomes comparable with the energy $\hbar\omega_0$
of the optical phonons and the Eliashberg theory based on the
adiabatic approximation $\varepsilon_F/\hbar\omega_0>>1$ breaks
down. Further, some authors have attempted to describe the
superconductivity in high-$T_c$ cuprates in terms of the BCS-BEC
crossover. According to the Landau criterion \cite{17}, BEC of an
ideal Bose gas of small real-space pairs and Cooper pairs is
irrelevant to the superconductivity (superfluidity) phenomenon.
Therefore, the superfluid transition in liquid $^4$He should not
be considered as the usual BEC of Bose particles. Because the
liquid $^4$He is strongly interacting Bose system and not an ideal
Bose gas which undergoes a BEC. Evans and Imry emphasized
\cite{18} that the superfluid phase in $^4He$ is best identified
with a nonvanishing coherence parameter of attracting bosons
rather than with the presence of BEC in ideal or repulsive Bose
gases where condensation can exist without coherence.

For a long time, the basic questions concerning the true origins
of the unusual superconducting/superfluid states in high-$T_c$
cuprates and other exotic systems remain open. In these systems,
unconventional interactions may take place between pairs of
quasiparticles, leading to new and unidentified states of matter.
Actually, the doped cuprates exhibit pseudogap phenomena \cite{10}
and diamagnetism \cite{19,20} above the superconducting transition
temperature $T_c$ and a $\lambda$-like superconducting transition
at $T_c$ \cite{21} just like the $\lambda$ transition in liquid
$^4$He. The pseudogap formation, diamagnetism, vortex-like
excitation, high-$T_c$ superconductivity and quantum criticality
occur in the cuprates between their insulating state at low doping
and their normal metallic state at high doping \cite{10,11,12,19}.
Attempts to understand the new physics of high-$T_c$
superconductors led to the assumptions of the importance of
superconducting fluctuations \cite{22,23}. It was argued that the
fluctuations of the BCS-like order parameter could be responsible
for the retaining of Cooper pairs and superconductivity on short
length scales (i.e., in small islands) at temperatures higher than
$T_c$. Further, the so-called $d$-wave superconductivity just like
the superconducting fluctuation is widely discussed in various
contexts and does not fundamentally differ from the BCS
superconductivity. But the validity of these scenarios for
superconductivity in high-$T_c$ cuprates is not justified. Because
both the superconducting fluctuation model and the $d$-wave
superconductivity model fails to account for the $\lambda$-like
transition and other unusual superconducting properties of the
cuprates. Remarkably, the cuprate compounds in the intermediate
doping regime exhibit unexplained exotic properties inherent in
unconventional superconductors (e.g., heavy-fermion systems
\cite{17,24}), quantum liquids ($^3$He and $^4$He)
\cite{17,25,26,27} and low-density nuclear matter \cite{28} but at
high doping levels they are similar to ordinary metals
\cite{10,29} and high-density nuclear matter. Apparently, the
essential physics of underdoped to overdoped cuprates,
heavy-fermion and organic superconductors, superfluid $^3$He,
superfluid atomic Fermi gases and superfluid low-density nuclear
matter may be described by a two-stage Fermi-Bose-liquid model
\cite{30,31} and controlled by the formation of bosonic Cooper
pairs and by the attractive interactions between these composite
bosons. This idea opens the way to consider that in more complex
pseudogap matters, the Bose-liquid superconductivity/superfluidity
might occur rather than the BCS-type Fermi-liquid
superconductivity/superfluidity. Another important fact is that
the superfluidity in ultracold atomic Fermi gases with an
extremely high transition temperature with respect to the Fermi
temperature $T_F\simeq5T_c$ defies also a BCS-like description
\cite{32}. The above arguments, together with the experimental
evidences for vortex-like state above $T_c$, two distinct
superconducting/superfluid phases below $T_c$, a $\lambda$-like
phase transition at $T_c$ and a first-order phase transition
somewhat below $T_c$ in high-$T_c$ cuprates and other
unconventional superconductors and superfluid $^3$He seemed to
make the BCS-like scenario as hopeless to explain fully the
unconventional superconductivity (superfluidity) and stimulated
the search for radically new mechanisms. At present, the greatest
part of the available experimental data is not accurate enough to
identify the superconduting order parameter in high-$T_c$ cuprates
and the different interpretations of the experimental results for
the pseudogap, vortex-like excitations and diamagnetism persisting
in the normal state of underdoped to overdoped cuprates are often
misleading. In particular, a prolonged dispute about the $s$-wave
or the $d$-wave superconducting gap in high-$T_c$ cuprates, which
is determined by using the single particle angle-resolved
photoemission spectroscopy (ARPES) and tunneling spectroscopy, is
already deadlocked. Thus, the properties of the pseudogap and
superconducting phases of these intricate materials are the
central issues in the search for the mechanism of high-$T_c$
cuprate superconductivity. In order to obtaine the novel types of
superconductivity and superfluidity in low Fermi energy systems,
there are two problems to be solved. First, the crossover from
fermionic limit of large Cooper pairs to bosonic limit of small
Cooper pairs is not well understood yet. Second, the BEC in ideal
Bose gases of small Cooper pairs exists without coherence. The
above interrelated two problems can be solved by considering the
real possibility of the bosonization of Cooper pairs and the
superfluid condensation of the attractive Bose gases of Cooper
pairs.

The present paper is devoted to discussion of these two important
questions. First we consider the possibility of formation of
bosonic Cooper pairs and formulate the criterion for bosonization
of Cooper pairs in high-$T_c$ cuprates and other pseudogap matters
by using the uncertainty principle. Then we consider the
condensation of the attractive Bose gases of Cooper pairs into a
superfluid Bose-liquid and the superfluidity of Bose-liquid with
coherent single particle and pair condensates, which arise from an
effective attraction between bosonic Cooper pairs in some domains
of momentum space, within the composite-boson mean field theory.
By solving the mean field equations for attractive Bose systems
and closely examining the possible superconducting/superfluid
states arising in high-$T_c$ cuprates and other classes of exotic
matters, we find that bosonic Cooper pairs and novel types of
superconductivity and superfluidity may indeed exist in such
systems. We then describe in detail the novel
superconducting/superfluid properties of these systems and their
experimental manifestations. We discuss the capabilities of the
existing experimental techniques for identifying the true
superconducting order parameter in high-$T_c$ cuprates. Further,
we describe the entire doping-temperature phase diagram of
high-$T_c$ cuprates from Mott insulator to the heavily overdoped
regime and the existence regions of the distinct superconducting
states below $T_c$ and the possible pseudogap, diamagnetic and
vortex-like states above $T_c$. The origins of vortices in
high-$T_c$ cuprates above $T_c$ and in thin $^4$He superfluid film
on porous substrate are explained naturally as the destruction of
the bulk superconductivity (superfluidity) and the remnant
quasi-two-dimensional (2D) superconductivity (superfluidity) above
$T_c$. We show that the superfluid Bose-liquid model provides a
fairly good quantitative description of unconventional
superconductivity (superfluidity) observed in high-$T_c$ cuprates,
heavy-fermion and organic compounds, quantum liquids ($^3$He and
$^4$He) and ultracold atomic Fermi gases. Finally, the basic
principles of novel superconductivity (superfluidity) in these
systems described by a two-stage Fermi-Bose-liquid model are
formulated.

\section{Criterion for bosonization of Cooper pairs}

There is now much experimental evidence that polaronic carriers
are present in doped cuprates \cite{33,34,35} and they have
effective masses $m_p\simeq(2-3)m_e$ \cite{29,33} (where $m_e$ is
the free electron mass) and binding energies $E_p \simeq
(0.06-0.12)$ eV \cite{34}. In lightly doped cuprates, polarons
tend to form real-space pairs, which are localized bipolarons. In
conventional metals, fermionic Cooper pairs and superconductivity
appear simultaneously at $T_c$. The situation, however, is
different in high-$T_c$ cuprates in which the electron-phonon
interactions are unconventional and the Fermi energy
$\varepsilon_F$ of polarons is comparable with the energy
$\hbar\omega_0$ of the high-frequency optical phonons. These
superconductors are characterized by low Fermi energies
$\varepsilon_F\simeq(0.1-0.3)$ eV \cite{36} and high-energy
optical phonons $\hbar\omega_0\simeq(0.04-0.08)$ eV \cite{33,36}.
Therefore, the Cooper pairing of polarons may occur in the normal
state of high-$T_c$ cuprates at a characteristic temperature $T^*$
\cite{31,37}. In these materials the attractive interaction
mechanism (e.g., due to exchange of static and dynamic phonons)
between the carriers operating in the energy range
$\{-(E_p+\hbar\omega_0),(E_p+\hbar\omega_0)\}$ is more effective
than in the simple BCS picture-in the narrow energy range
$\{-\hbar\omega_D,\hbar\omega_D\}$. For such strong pairing
interactions, it is predicted that the pseudogap phase has a
BCS-like dispersion given by
$E({k})=\sqrt{\varepsilon_F^2({k})+\Delta_F^2}$ (where
$\varepsilon_F({k})$ is the energy of fermionic quasiparticles
measured from the Fermi energy $\varepsilon_F$, ${k}$ is the
quasiparticle momentum), but the BCS-like gap $\Delta_F$ is no
longer the superconducting order parameter and opens on the Fermi
surface at $T^*>T_c$ \cite{31}. Various experiments showed that a
BCS-like excitation gap indeed persists as a pseudogap well above
the measured critical temperature for
superconductivity/superfluidity in high-$T_c$ cuprates and atomic
Fermi gases \cite{32}. In particular, the formation of such a
pseudogap at the precursor Cooper pairing of polarons with
antiparallel spins is manifested in the diamagnetic property of
high-$T_c$ cuprates above $T_c$.

As the binding between fermions increases, Fermi gas of
weakly-bound Cooper pairs evolves into Bose gas of tightly-bound
Cooper pairs, as pointed out by Leggett \cite{15}. This is the
most interesting crossover regime, since a Fermi system passes
from a BCS-like Fermi-liquid limit to a normal Bose gas limit with
decreasing $\varepsilon_F$. Thus, it is a challenging problem to
find the criterion for bosonization of Cooper pairs in such Fermi
systems. If the size of the Cooper pairs $a_c(T)$ is much larger
than the average distance $R_c$ between them, the bosonization of
such Cooper pairs cannot be realized due to their strong
overlapping, as argued by Bardeen and Schrieffer \cite{38,39}.
However, the composite (bosonic) nature of Cooper pairs becomes
apparent when $a_{c}\sim R_{c}$. At $R_c\gtrsim a_{c}$, the
fermions cannot move from one Cooper pair to another one and the
non-overlapping Cooper pairs behave like bosons. The criterion for
bosonization of polaronic Cooper pairs can be determined from the
uncertainty relation \cite{40}
\begin{eqnarray}\label{Eq.1}
\Delta x\cdot\Delta E\simeq\frac{(\hbar\Delta
k)^2}{2m_p}\frac{1}{2\Delta k}
\end{eqnarray}
where $\Delta x$ and $\Delta E$ are the uncertainties in the
coordinate and energy of attracting polaronic carriers, $\Delta k$
is the uncertainty in the wave vector of polarons. The expression
$(\hbar\Delta k)^2/2m_p$ represents the uncertainty in the kinetic
energy of polarons, which is of order $\varepsilon_F$, whereas
$\Delta k$ would be of the order of $1/R_c$. Taking into account
that $\Delta x$ is of order $a_c$ and $\Delta E$ would be of the
order of the characteristic energy $\varepsilon_A$ of the attractive
interaction between polarons, Eq. (\ref{Eq.1}) can be written as
\begin{eqnarray}\label{Eq.2}
\frac{R_c}{a_c}\simeq2\frac{\varepsilon_A}{\varepsilon_F}\gtrsim1
\end{eqnarray}
This ratio is universal criterion for the bosonization of Cooper
pairs in low Fermi energy systems, in particular, in high-$T_c$
cuprates (for which $\varepsilon_A$ is replaced by
$E_p+\hbar\omega_0$) and other exotic superconductors, liquid
$^3$He and ultracold atomic Fermi gases. The criterion in Eq.
(\ref{Eq.2}) is well satisfied at
$\varepsilon_A\gtrsim0.5\varepsilon_F$, where
$\varepsilon_F\simeq0.016-0.025$ eV (for UPt$_3$ \cite{41}),
$\varepsilon_F\simeq0.1-0.3$ eV (for organic compounds \cite{36}),
 $\varepsilon_F\simeq4.4\times10^{-4}$ eV (for liquid $^3$He
\cite{42}) and $\varepsilon_F\simeq10^{-10}$ eV (for ultracold
atomic Fermi gases \cite{43}). For the mass density of nuclear
matter $\rho_M\simeq10^{11}$ g/cm$^3$ \cite{44}, we find
$\varepsilon_F\simeq0.38$ MeV. Then the deuteron-like bosonic
Cooper pairs in low-density nuclear matter are formed at
$\varepsilon_A>0.19$ MeV. We can now formulate the following key
postulates:

(1) the BCS-type superconductivity and superfluidity would occur
in the fermionic limit of Cooper pairs and could exist in
high-density nuclear matter and conventional superconductors
(e.g., ordinary metals and heavily overdoped cuprates) with large
Fermi energies, where the pseudogap is absent and the
superconducting state is characterized by the BCS-like order
parameter and the onset temperature of Cooper pairing $T^*$
coincides with $T_c$;

(2) the high-$T_c$ cuprates exhibiting pseudogap behaviors at
$\varepsilon_F<<1$ eV and other pseudogap matters (e.g.,
heavy-fermion and organic compounds, liquid $^3$He, atomic Fermi
gases and low-density nuclear matter) could be in the bosonic
limit of Cooper pairs and the novel (non-BCS-type)
superconductivity and superfluidity would occur in such systems,
where the pseudogap coexists with the unusual superconducting
order parameter below $T_c$ \cite{31}. In pseudogap matters, a
two-stage superconducting/superfluid transition process involves
the formation of bosonic Cooper pairs at $T^*>T_c$ and the
subsequent superfluid Bose condensation of these bosons at $T_c$.

\section{Composite-boson mean field theory of superfluidity and its experimental confirmation}

The repulsive interaction is less realistic for the problem of
Bose superfluids (including $^4$He). In unconventional
superconductors/superfluids with
$\varepsilon_F\lesssim2\varepsilon_A$, Cooper pairs behave as
composite bosons and would undergo a BEC in the noninteracting
particle approximation without superfluidity at $T=T_{BEC}>>T_c$.
Here we show that the superconductivity/superfluidity in these
systems is driven by the condensation of the attractive Bose gases
of Cooper pairs with low densities. Such composite bosons repel
one another at small distances between them and their net
interaction is attractive at large distances. The Hamiltonian of a
Bose gas interacting via a pair potential $V_B(k-k')$ has the form
\cite{45}
\begin{eqnarray}\label{Eq.3}
H_B=\sum_k[\tilde{\varepsilon}_B(k)c^+_{k}c_k+\frac{1}{2}\Delta_B(k)(c^+_{-k}c^+_k+c_{k}c_{-k})],
\end{eqnarray}
where
$\tilde{\varepsilon}_B(k)=\varepsilon(k)-\mu_B+V_B(0)\rho_B+\chi_B(k)$,
$\varepsilon(k)=\hbar^2k^2/2m_B$, $\mu_B$ is the chemical potential,
$\Delta_B(k)=(1/\Omega)\sum_{k'}V_B(k-k')<c_{-k'}c_{k'}>$ is the
coherence parameter, $\chi_B(k)=(1/\Omega)\sum_{k'}V_B(k-k')n_B(k)$,
$n_B(k)=<c^+_{k}c_k>$ is the particle number operator,
$\rho_B=(1/\Omega)\sum_{k'}n_B(k')$, $m_B=2m_p$ is the mass of
bosonic Cooper pairs, $c^+_{k}(c_k)$ is the creation (annihilation)
operator of bosons with the wave vector $k$, $\Omega$ is the volume
of the system.

The Hamiltonian in Eq. (\ref{Eq.3}) is diagonalized by the
Bogoliubov transformations of Bose operators and the quasiparticle
spectrum has the form
$E_B(k)=\sqrt{\tilde{\varepsilon}_B^2(k)-\Delta^2_B(k)}$, which is
gapless for $k=0$ and $k'=0$ provided
$\tilde{\mu}_B=-\mu_B+V_B(0)\rho_B+\chi_B(0)=|\Delta_B(0)|$. If
$E_B(k)=0$, the $k=0$ and $k'=0$ terms in the summation of the
equations for $\Delta_B(k)$, $\chi_B(k)$ and $\rho_B$ are
considered separately according to the procedure proposed in Ref.
\cite{18}. Further, in order to simplify the solutions of the
equations for $\Delta_B(k)$, $\chi_B(k)$ and $\rho_B$, the
interboson interaction potential may be chosen in a simple
separable form
\begin{eqnarray}\label{Eq.4}
\hspace{-0.5cm}V_B(k-k')=\left\{\begin{array}{lll}
V_{BR}-V_{BA}\\
V_{BR}\\
0\\
\end{array}
\begin{array}{lll}
\textrm{for}\: \ \varepsilon(k), \ \varepsilon(k')<\varepsilon_{BA}, \ \\
\hspace{-0.9cm}\textrm{for}\: \varepsilon_{BA}\leq \varepsilon(k), \ \ \varepsilon(k')<\varepsilon_{BR}, \ \\
\hspace{-0.9cm}\textrm{for}\: \varepsilon(k), \
\varepsilon(k')>\varepsilon_{BR},
\end{array}
\right.
\end{eqnarray}
where $\varepsilon_{BA}$ and $\varepsilon_{BR}$ are the cutoff
parameters for attractive $V_{BA}$ and repulsive $V_{BR}$ parts of
$V_B(k-k')$, respectively. Then the three-dimensional (3D) equations
for determining the coherence (e.g., superconducting order)
parameter $\Delta_{SC}=\Delta_B$ and the condensation temperature
$T_c$ of attracting bosons can be written as:
\begin{eqnarray}\label{Eq.5}
\frac{2}{D_B\tilde{V}_B}=\int^{\varepsilon_{BA}}_0
\sqrt{\varepsilon}\frac{\coth\left[\frac{\sqrt{(\varepsilon+\tilde{\mu}_B)^2-\Delta^2_B}}{2k_BT}\right]}{\sqrt{(\varepsilon+\tilde{\mu}_B)^2-\Delta^2_B}}d\varepsilon,
\end{eqnarray}
\begin{eqnarray}\label{Eq.6}
\frac{2\rho_B}{D_B}&=&\int^\infty_0
\sqrt{\varepsilon}\Bigg\{\frac{\varepsilon+\tilde{\mu}_B}{\sqrt{(\varepsilon+\tilde{\mu}_B)^2-\Delta^2_B}}\times
\nonumber\\&& \times
\coth\left[\frac{\sqrt{(\varepsilon+\tilde{\mu}_B)^2-\Delta^2_B}}{2k_BT}\right]-1\Bigg\}d\varepsilon,
\end{eqnarray}
where $D_B=m^{3/2}_B/\sqrt{2}\pi^2\hbar^3$,
$\tilde{V}_B=V_{BA}-V_{BR}[1+V_{BR}I_R]^{-1}$, $I_R\simeq
D_B[\sqrt{\varepsilon_{BR}}-\sqrt{\varepsilon_{BA}}]$,
$\varepsilon_{BA}<<\varepsilon_{BR}$.

Solutions of Eqs. (\ref{Eq.5}) and (\ref{Eq.6}) allow us to
examine closely the possible superconducting/superfluid states
arising in attractive 3D Bose systems. Below $T_c$, the excitation
spectrum $E_B(k)$ has a gap
$\Delta_g=\sqrt{\tilde{\mu}_B-\Delta^2_{B}}$ and satisfies the
Landau criterion for superfluidity. For
$\gamma_B=D_B\tilde{V}_B\sqrt{\varepsilon_{BA}}$ less than a
threshold value $\gamma_{B}^*$, however, $E_B(k)$ becomes gapless
at $T\leq T^*_{c}<<T_c$ (for $\gamma_B<\gamma_B^*$) or at $T\leq
T^*_{c}<T_c$ (for $\gamma_B<<1$). For $T=0$ and
$\varepsilon_{BA}/k_BT_{BEC}=10-50$, the energy gap $\Delta_g$
vanishes at the critical values of the interboson coupling
constant $\gamma_B=\gamma_B^*\simeq1.4-2.0$. When the interboson
coupling is weak ($\gamma_B<<1$), the coherence parameter
$\Delta_B$ is proportional to the density of condensed bosons
$\rho_B$ (i.e., $\Delta_B\simeq\rho_B\tilde{V_B}$) \cite{46,47}.
For $\gamma_B<\gamma^*_B$ and $\Delta_g=0$, Eqs. (\ref{Eq.5}) and
(\ref{Eq.6}) become
\begin{eqnarray}\label{Eq.7}
\frac{2}{D_B\tilde{V}_B}=\frac{2\rho_{B0}}{D_B\tilde{\mu}_B}+\int^{\varepsilon_{BA}}_0
\sqrt{\varepsilon}\frac{\coth\left[\frac{\sqrt{\varepsilon^2+2\tilde{\mu}_B\varepsilon}}{2k_BT}\right]}{\sqrt{\varepsilon^2+2\tilde{\mu}_B\varepsilon}}d\varepsilon,
\end{eqnarray}
\begin{eqnarray}\label{Eq.8}
\frac{2\rho_B}{D_B}&=&\frac{2\rho_{B0}}{D_B}+\int^\infty_0
\sqrt{\varepsilon}\Bigg\{\frac{\varepsilon+\tilde{\mu}_B}{\sqrt{\varepsilon^2+2\tilde{\mu}_B\varepsilon}}\times
\nonumber\\&& \times
\coth\left[\frac{\sqrt{\varepsilon^2+2\tilde{\mu}_B\varepsilon}}{2k_BT}\right]-1\Bigg\}d\varepsilon,
\end{eqnarray}
where $\rho_{B0}$ is the density of condensed bosons with $k=0$ and
$\varepsilon=0$.

Equations (5) and (6) similar to those of the BCS theory for
fermions have collective solutions for the attractive interboson
interaction $\tilde{V}_B$. The superconducting/superfluid state is
characterized by the coherence (macroscopic order) parameter
$\Delta_B$ which vanishes at $T=T_c$, that marks the vanishing of
a macroscopic superfluid condensate of attracting bosons. For
$T\leq T_c^*$, the gapless and linear (at small $k$), phonon-like
spectrum $E_B(k)$ in the superfluid state is similar to the
excitation spectrum in superfluid $^4$He and satisfies also the
criterion for superfluidity, i.e., the critical velocity of
quasiparticles $v_c=\hbar^{-1}{(\partial E_B(k)/\partial
{k})}_{min}>0$ satisfies the condition for the existence of
superfluidity. By solving the mean field equations (5) and (6) for
$\Delta_{g}>0$ and $\Delta_{g}=0$, we find that the condensation
of boson pairs at $T>T^*_c$ will correspond to a smaller value of
both the chemical potential $\tilde{\mu}_B$ and the order
parameter $\Delta_B<\tilde{\mu}_B$, while the single particle
condensation of bosons at $T\leq T^*_c$ will correspond to a much
larger value of the chemical potential $\tilde{\mu}_B=\Delta_B$.
In this case, the pair condensation of attracting bosons occurs
first at $T_c$ and their single particle condensation takes place
at $T^*_c <T_c$ (for $\gamma_B<<1$) or at $T^*_c <<T_c$ (for
$\gamma_B\gtrsim1$). First the $\lambda$-like second-order phase
transition in high-$T_c$ cuprates occurs at $T_c$. Then, a new
first-order phase transition from pair condensation state to
single particle condensation one occurs at $T\leq T^*_c$ and the
true superconducting order parameter $\Delta_{SC}(T)$ shows a
pronounced kink-like behavior near $T^*_c$. We see that two
distinct superconducting states of a 3D attractive Bose gas of
Cooper pairs in high-$T_c$ cuprates are characterized by the
integer $h/2e$ (at $T\leq T^*_c$ and
$\tilde{\mu}_B(T)=\Delta_B(T)$) and half-integer $h/4e$ (at
$T>T^*_c$ and $\tilde{\mu}_B(T)>\Delta_B(T)$) magnetic flux
quantizations. The first-order phase transition was actually
observed in high-$T_c$ cuprates \cite{48,49} and in superfluid
$^3$He (where the transition between the $A$ and $B$ phases occurs
at $T^*_c=T_{AB}$) \cite{26,50}. A similar phase transition was
also observed in heavy-fermion systems \cite{17,41,51}. Some
experiments \cite{52} indicate that the superconducting order
parameter $\Delta_{SC}(T)$ in the cuprates has a kink-like feature
near the characteristic temperature $T^*_c$ ($\lesssim0.6T_c$).
The half-integer circulation quantum $h/2m_4$ (where $m_4$ is the
mass of $^4$He atoms) observed in superfluid $^4$He \cite{53} and
the formation of $\alpha$-clusters in exotic nuclear matter
\cite{54} are equally well explained by the pair condensation of
$^4$He atoms and deuteron-like Cooper pairs. The half-quantum
vortices in the superfluid $^3$He-A were discussed in Ref.
\cite{55}. The microscopic origins of the half-quantum vortices
($h/4e$ and $h/4m_3$) in high-$T_c$ cuprates and superfluid
$^3$He-A are associated with the excitations of pair condensates
of bosonic Cooper pairs rather than other effects. The
superfluidity of $^3$He-B is caused by the single particle
condensation of attracting bosonic Cooper pairs.

Thus, the single particle and pair condensates of bosonic Cooper
pairs are different superconducting/superfluid phases in
high-$T_c$ cuprates and other pseudogap matters. The occurrence of
novel superconductivity/superfluidity in these systems is
characterized by a non-zero coherence parameter $\Delta_B$ which
defines the bond strength of all condensed bosons - boson
superfluid stiffness. Therefore, excitations of a superfluid Bose
condensate of Cooper pairs in high-$T_c$ cuprates are really
many-particle ones and cannot be measured by single-particle
spectroscopies, as noted also in Ref. \cite{56}. In these systems
the gapless superconductivity/superfluidity occurs due to the
vanishing of the gap $\Delta_g$ in $E_B(k)$ at $T\leq T_c^*$ and
is not associated with the point or line nodes of the BCS-like gap
assumed in some $p$- and $d$-wave pairing models. The frictionless
flow of Bose condensate would be possible under the condition
$\Delta_B>0$. While the BCS-like fermionic excitation gap
$\Delta_{F}$ characterizing the bond strength of Cooper pairs may
exist as the pseudogap and its formation is not accompanied by the
superconducting transition \cite{57,58}. The other key
superconducting/superfluid properties of high-$T_c$ cuprates and
related systems will be discussed below.

\section{Experimental manifestations of novel superconducting and superfluid properties}

\begin{figure}[!htp]
\begin{center}
\includegraphics[width=0.45\textwidth]{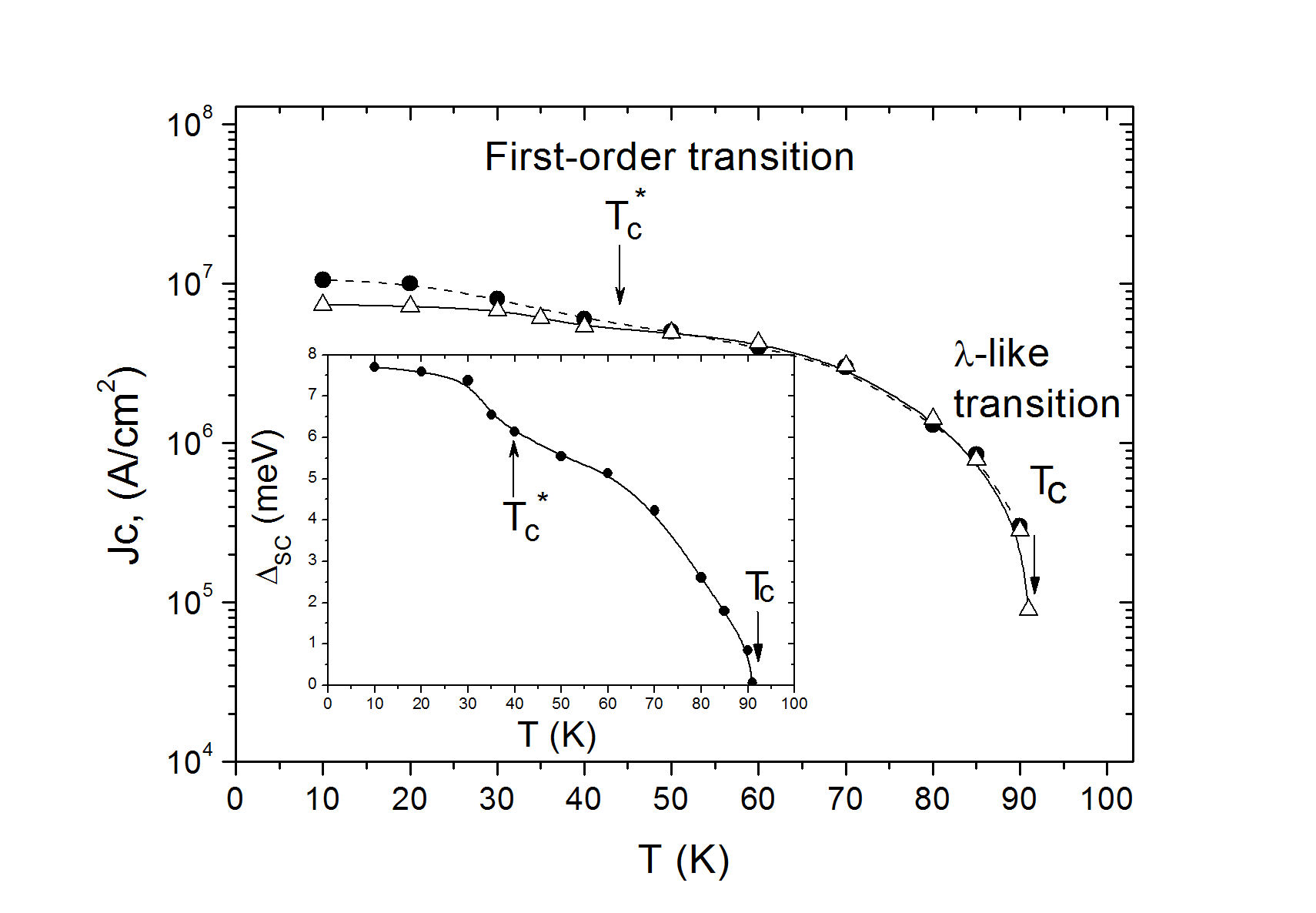}
\caption{\label{fig.1} Temperature dependence of $J_c$ measured in
YBCO film ($\bullet$). The solid line is the best fit of Eq.
(\ref{Eq.9}) ($\triangle$) to the experimental data ($\bullet$)
for YBCO film \cite{59} using the parameters
$\rho_B\simeq0.8\times10^{19} cm^{-3}$, $m_B=4.6m_e$ and
$\varepsilon_{BA}=0.08$ eV. The inset shows the kink-like behavior
of $\Delta_{SC}(T)$ near $T^*_c$.}
\end{center}
\end{figure}

\begin{figure}[!htp]
\begin{center}
\includegraphics[width=0.45\textwidth]{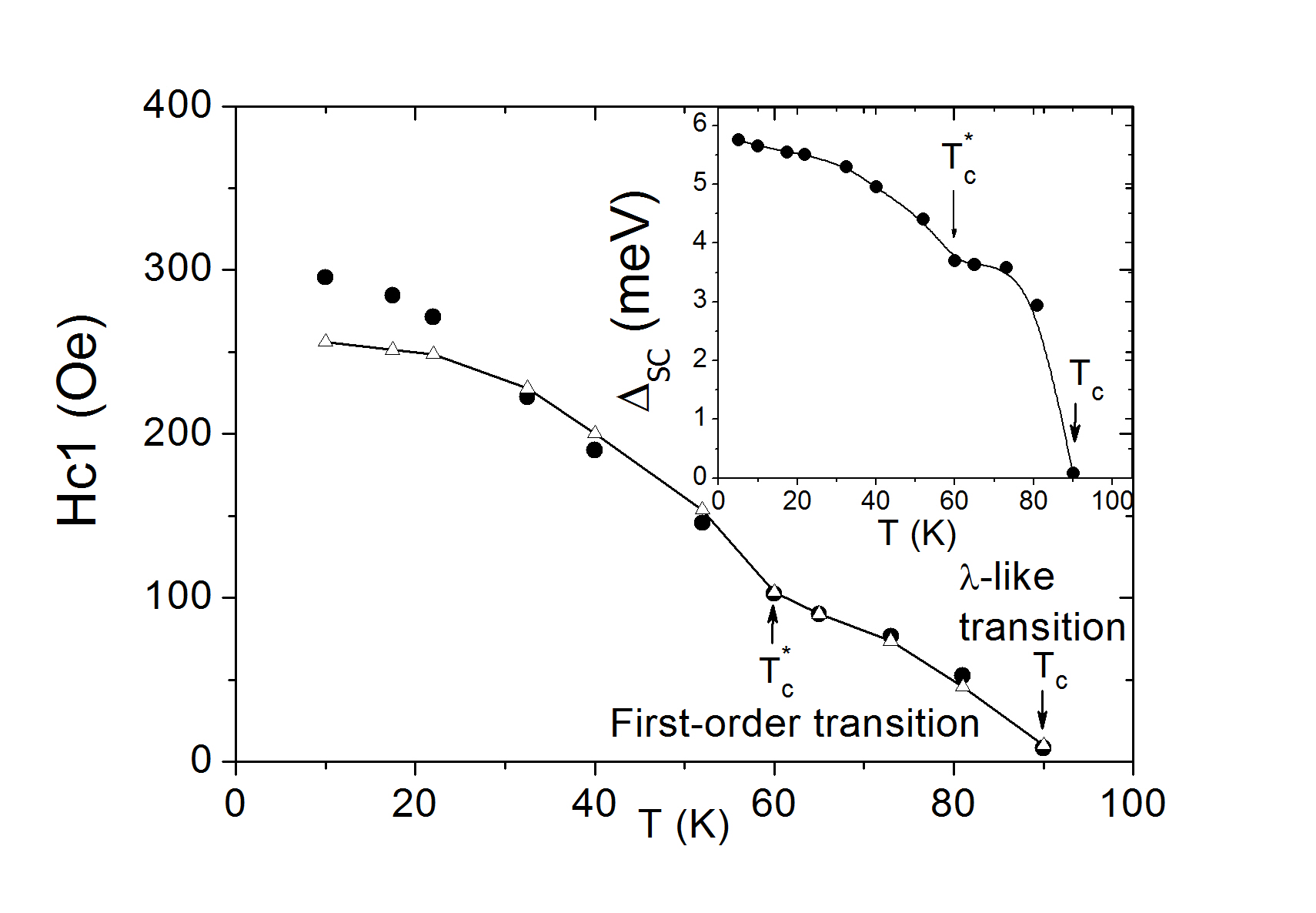}
\includegraphics[width=0.45\textwidth]{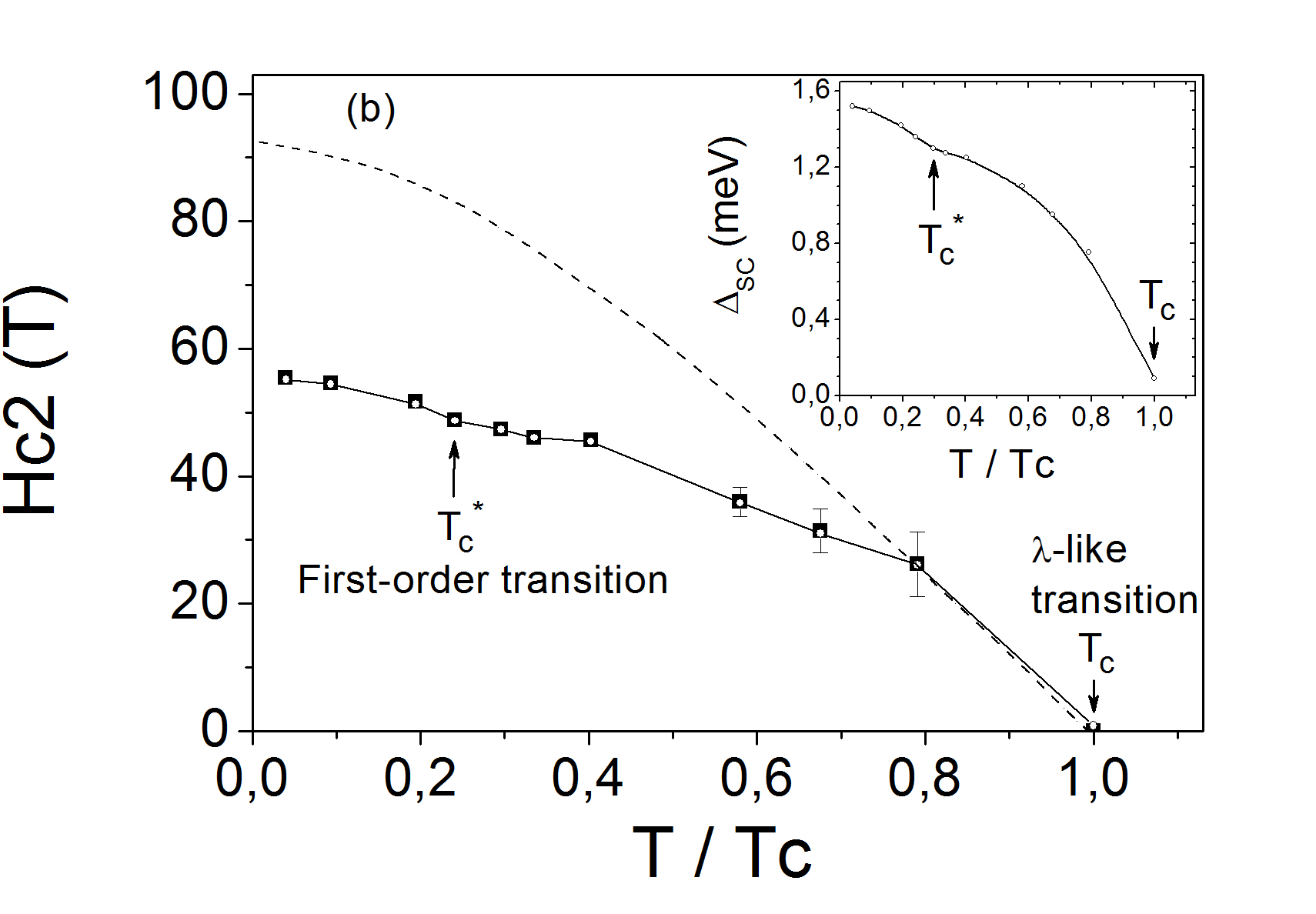}
\caption{\label{fig.2} Temperature dependences of the critical
magnetic fields $H_{c1}(T)$ and $H_{c2}(T)$ measured in
superconducting cuprates. (a) Solid line is the fit of Eq.
(\ref{Eq.10}) ($\triangle$) to the experimental data ($\bullet$)
for $H_{c1}(T)$ in YBCO \cite{60} using the parameters
$\rho_B\simeq1.7\times10^{19} cm^{-3}$, $m_B=4.4m_e$, $R=0.01 cm$
and $\varepsilon_{BA}=0.18$ eV. (b) Solid line is the fit of
equation $H_{c2}(T)=\sqrt{2}\chi(T)H_{c}(T)$ ($\circ$) to the
experimental data ($\blacksquare$) for $H_{c2}(T)$ in Bi-2201
\cite{61} using the parameters $\rho_B\simeq0.1\times10^{19}
cm^{-3}$, $m_B=5m_e$, $R=0.5\times10^{-3} cm$ and
$\varepsilon_{BA}=0.13$ eV. Dashed line is by the
Werthamer-Helfand-Hohenberg theory (see Ref. \cite{61}). Insets
show the kink-like behaviors of $\Delta_{SC}(T)$ near $T^*_c$.}
\end{center}
\end{figure}

We now discuss the novel superconducting properties and show that
the kink-like features of $\Delta_{SC}(T)$ are responsible for the
kink-like behaviors of the critical current $J_c(T)$ and the
critical magnetic fields ($H_{c1}(T)$ and $H_{c2}(T)$) near
$T^*_c$, as observed in various high-$T_c$ cuprates
\cite{59,60,61}. The critical current density can be written as
\begin{eqnarray}\label{Eq.9}
J_c(T)=2e\rho_s(T)v_c(T),
\end{eqnarray}
where $\rho_s(T)=\rho_B-\rho_n$ is the density of the superfluid
part of condensed bosons, $\rho_n=-(1/3m_B)\int(dn_B/dE_B)p^2[4\pi
p^2/(2\pi\hbar)^3]dp$ is the density of the normal part of a 3D
Bose-liquid, $p=\sqrt{2m_B\varepsilon}$,
$n_B=[exp(E_B(k)/k_BT)-1]^{-1}$,
$v_c(T)=\sqrt{[\tilde{\mu}_B(T)+\Delta_g(T)]/m_B}$ is the critical
velocity of superfluid carriers. The kink-like behavior of
$J_c(T)$ in $\rm{YBa_2Cu_3O_7}$ (YBCO) film is shown in Fig.
\ref{fig.1}.
The lower critical magnetic field $H_{c1}$ is determined from the
relation
\begin{eqnarray}\label{Eq.10}
H_{c1}(T)=\frac{\ln\chi(T)}{\sqrt{2}\chi(T)}H_c(T),
\end{eqnarray}
where $\chi(T)=\lambda_L(T)/\xi_c(T)$ is the Ginzburg-Landau
parameter, $\lambda_L(T)=[m_Bc^2/16\pi e^2\rho_s(T)]^{1/2}$ is the
London penetration depth, $\xi_c(T)=\hbar/\sqrt{2m_B\Delta_B(T)}$
is the coherence length of bosonic superconductors, $H_c(T)=4\pi
RJ_c(T)/c$ is the thermodynamic critical magnetic field, $R$ is
the radius of a superconducting wire, $c$ is the velocity of
light. The kink-like behaviors of $H_{c1}(T)$ (in YBCO) and upper
critical magnetic field $H_{c2}(T)$ (in
$\rm{Bi_{2+x}Sr_{2-x}CuO_6}$ (Bi-2201) with $T_c\lesssim15$ K)
near $T_c^*$ are shown in Fig. \ref{fig.2}. A peak in the specific
heat of high-$T_c$ cuprates \cite{62} and heavy-fermion compounds
\cite{17,41,51} was also observed at $T_c^*$ below which
$H_{c1}(T)$ suddenly increased. Further, an abrupt jump-like
increase of the critical velocity $v_c(T)=\sqrt{\Delta_B(T)/m_B}$
three times and such a change of the superfluid density at
$T\simeq(0.6-0.7)T_c$ were observed in superfluid $^3$He
\cite{55}. Clearly, the sharp increasing of $\Delta_B(T)$ at the
vanishing of the gap $\Delta_g$ in $E_B(k)$ near $T_c^*$ leads to
the jump-like increasing of $v_c(T)$ and superfluid density
$\rho_s(T)$ at $T\leq T_c^*$.

The specific heat of a 3D superfluid Bose-liquid $C_v(T)$,
diverges as $C_v(T)\sim (T_c-T)^{-0.5}$ near $T_c$ (where
$\Delta_B(T)<<\tilde{\mu}_B(T)<<k_BT_c$) \cite{46,47} and will
exhibit a $\lambda$-like anomaly at $T_c$, as observed in
high-$T_c$ cuprates \cite{21,27}, organic superconductors
\cite{17} and superfluid $^3$He (see Fig. 1.9a in Ref. \cite{42}).
Such a behavior of $C_v(T)$ is similar to that of superfluid
$^4$He. Note that, as $T$ approaches $T_c$ from below, the
temperature dependences of $\tilde{\mu}_B$ and $\Delta_B$ are
defined as
\begin{eqnarray}\label{Eq.11}
\tilde{\mu}_B(T)\simeq\tilde{\mu}_B(T_c)\left[1+a(T_c-T)^{0.5}\right]
\end{eqnarray}
and
\begin{eqnarray}\label{Eq.12}
\Delta_B(T)\simeq2\tilde{\mu}_B(T_c)\sqrt{a}(T_c-T)^{0.25},
\end{eqnarray}
where $a=2(c_0\gamma_BT_c)^{-0.5}(\varepsilon_{BA}/k_BT_c)^{0.25}$
and $c_0=\pi^{3/2}/3.912$. Therefore, the temperature derivatives
of $\tilde{\mu}_B$ and $\Delta_B$ entering the expression for
$C_v(T)$ give rise to a pronounced $\lambda$-like divergence. By
introducing the quantity of superfluid matter $\nu_B=N_B/N_A$
(where $N_B$ is the number of attracting bosonic Cooper pairs and
$N_A$ is the Avogadro number, which is equal to the number of
$CuO_2$ formula unit per unit molar volume) and the molar fraction
of the superfluid bosonic carriers defined by $f_s=\nu_B/\nu$
(where $\nu=N/N_A$ is the amount of doped matter), we now write
the molar specific heat of the superfluid Bose-gas in high-$T_c$
cuprates as
\begin{eqnarray}\label{Eq.13}
\hspace{-0.7cm}C_s(T)&\hspace{-0.7cm}=\hspace{-0.7cm}&f_s\frac{C_v(T)}{\nu_B}=f_s\frac{D_Bk_BN_A}{4\rho_B(k_BT)^2}
\int^{\varepsilon_{BA}}_0\hspace{-0.3cm}\sqrt{\varepsilon}
\frac{d\varepsilon}{\sinh^2\frac{E_B(\varepsilon)}{k_BT}}\times\nonumber\\&&
\times\left\{E^2_B(\varepsilon)+\frac{a\tilde{\mu}_B(T_c)T}{2(T_c-T)^{0.5}}
\left[\varepsilon-\tilde{\mu}_B(T_c)\right]\right\}.
\end{eqnarray}
Here we have accounted for that $\Omega/\nu_B=N_Bv_B/\nu_B=v_BN_A$
and $v_B=1/\rho_B$. In doped cuprates the carriers are distributed
between the polaronic band and the impurity band (with Fermi
energy $\varepsilon_{FI}$) and the normal-state specific heat
$C_n(T)$ above $T_c$ is calculated by considering three
contributions from the excited components of Cooper pairs, the
ideal Bose-gas of Cooper pairs and the unpaired carriers bound to
impurities \cite{63}.
\begin{figure}[!htp]
\begin{center}
\includegraphics[width=0.45\textwidth]{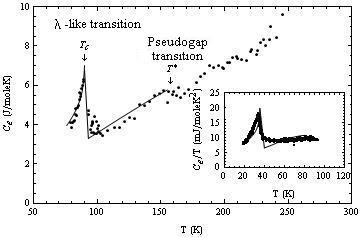}
\caption{\label{fig.3} Temperature dependence of the specific heat
of $HoBa_2Cu_3O_{7-\delta}$ measured near $T_c$ and above $T_c$
\cite{64}. Solid line is the calculated curve for comparing with
experimental points (black circles). According to \cite{63},
$C_n(T)$ is calculated by using the parameters
$\varepsilon_F=0.12$ eV, $\varepsilon_{FI}=0.012$ eV, $f_p=0.3$,
$f_I=0.7$, while superconducting contribution $C_s(T)$ to $C_e(T)$
is calculated by using the parameters $\rho_B=1.6\times10^{19}
cm^{-3}$, $m_B=2.5m_p$, $\tilde{\mu}_B(T_c)=1.6$ meV and
$f_s=0.03$. The inset shows the calculated temperature dependence
of $C_e(T)/T$ (solid line) compared with experimental $C_e(T)/T$
data for LSCO \cite{27} (black circles). According to \cite{63},
$C_n(T)/T$ is calculated by using the parameters
$\varepsilon_{F}=0.1$ eV $\varepsilon_{FI}=0.06$ eV, $f_p=0.4$,
$f_I=0.6$, while $C_s(T)/T$ is calculated by using the parameters
$\rho_B=1.4\times10^{19} cm^{-3}$, $m_B=2.7m_p$,
$\tilde{\mu}_B(T_c)=0.5$ meV and $f_s=0.012$.}
\end{center}
\end{figure}
The fraction $f_p$ of carriers residing in the polaronic band and
the other fraction $f_I$ of carriers residing in the impurity band
are taken into account in comparing the specific heat $C_s(T)$
with the experiment. The total electronic specific heat
$C_e(T)=C_s(T)+C_n(T)$ below $T_c$ is calculated and compared with
the experimental data for $C_e(T)$ in cuprates (Fig. \ref{fig.3}).
The calculated results for $(\lambda_L(0)/\lambda_L(T))^2$ are
also compared with the experimental data (Fig. \ref{fig.4}).
\begin{figure}[!htp]
\begin{center}
\includegraphics[width=0.45\textwidth]{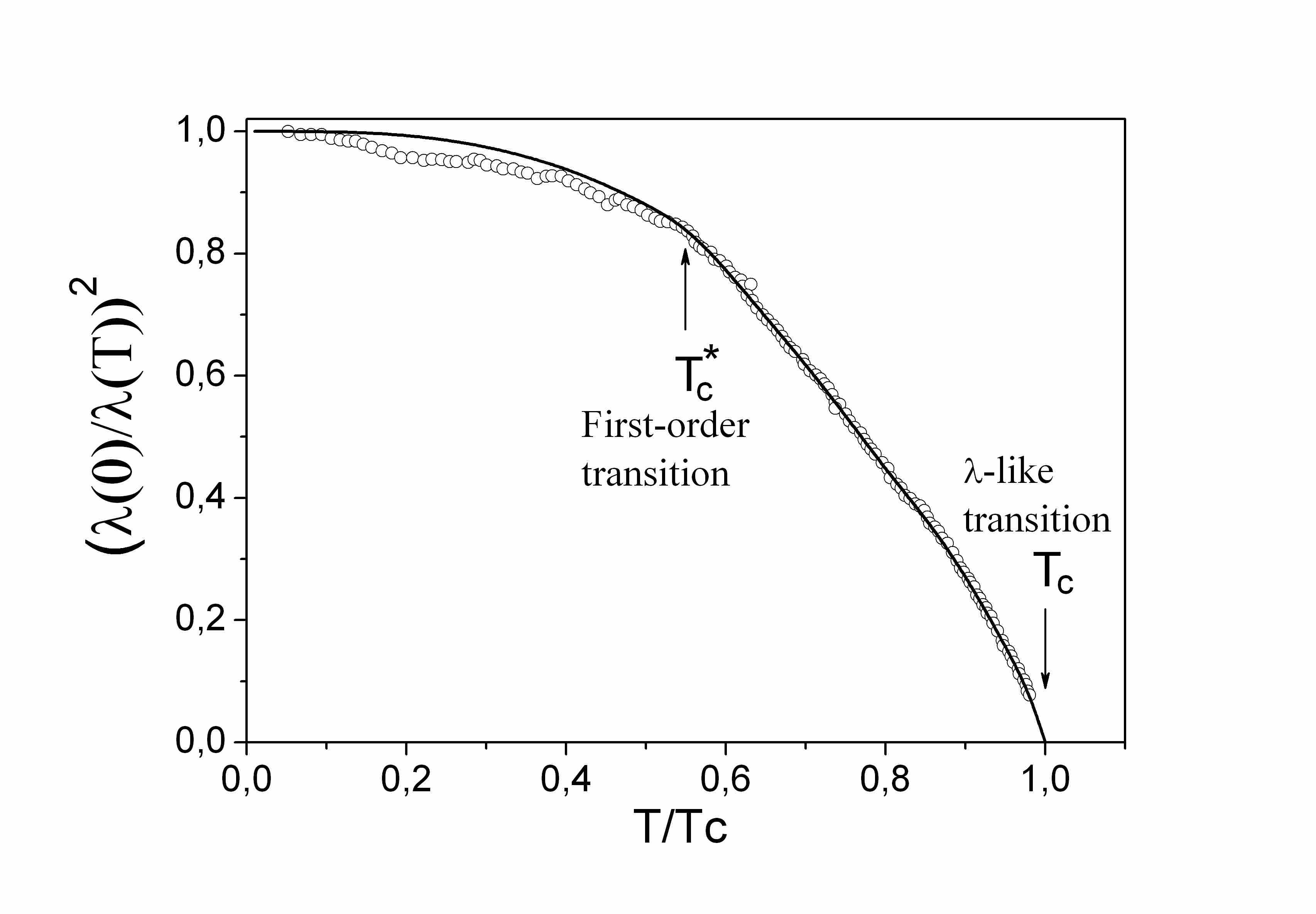}
\caption{\label{fig.4} Temperature dependence of
$(\lambda_L(0)/\lambda_L(T))^2$ (solid line) is calculated by
using the parameters $\rho_B\simeq1.29\times10^{19} cm^{-3}$,
$m_B=5m_e$ and $\varepsilon_{BA}=0.08$ eV and compared with
experimental data ($\circ$) for YBCO film \cite{65}}
\end{center}
\end{figure}

Analytical solutions of Eqs. (\ref{Eq.5}) and (\ref{Eq.6}) near
$T_c$ \cite{45,46} allow to estimate the bulk
superconducting/superfluid transition temperature as
\begin{eqnarray}\label{Eq.14}
T_c=T^{3D}_c\simeq
T_{BEC}\left[1+c_0\gamma_B\sqrt{\frac{k_BT_{BEC}}{\varepsilon_{BA}}}\right],
\end{eqnarray}
where $T_{BEC}=3.31\hbar^2\rho^{2/3}_B/k_Bm_B$, $\gamma_B<<1$.

By solving now the mean field equations for attractive 2D Bose
systems, we see that the pair condensation of bosons will take
place in the temperature range $0<T\leq T_c$, while their single
particle condensation occurs only at $T=0$. The condensation
temperature of attracting 2D bosons for arbitrary
$\gamma_B=D_B\tilde{V}_B$ is given by \cite{47}
\begin{eqnarray}\label{Eq.15}
T^{2D}_c=-\frac{T_0}{ln\left[1-\exp\left(-\frac{2\gamma_B}{2+\gamma_B}\right)\right]},
\end{eqnarray}
which is in an agreement with the result of Ref. \cite{66} at
$\gamma_B<<1$, where $T_0= 2\pi\hbar^2\rho_B/k_Bm_B$ and
$\tilde{V}_B$ depends now on
$I_R=D_{B}ln(\varepsilon_{BR}/\varepsilon_{BA}$) and
$D_B=m_B/2\pi\hbar^2$. When the effective (renormalized) mass of
interacting bosons depends on $\rho_B$, the mass of bosons $m_B$
in the expressions for $T_{BEC}$ and $T_0$ should be replaced by
$m^*_B=m_B[1-\rho_BV_B(0)/\varepsilon_{BR}]^{-1}$ \cite{47}. Thus,
both the $T_c$ and the $T^{2D}_c$ is mainly controlled by $\rho_B$
and $\gamma_B$.

\section{Vortex-like excitations above $T_c$ and gapless superconductivity and superfluidity below $T_c$}

It would also be interesting to discuss the conditions for (i) the
diamagnetism in the pseudogap state and vortex formation observed
above the bulk superconducting transition temperature $T_c$ in
high-$T_c$ cuprates, (ii) the vortex formation observed above the
bulk superfluid transition temperature $T_{\lambda}$ in liquid
$^4$He, and (iii) the gapless superconductivity and superfluidity
observed in high-$T_c$ cuprates and other systems somewhat below
$T_c$ or far below $T_c$. We first discuss the origins of the
gapless excitations and other effects observed in the
superconducting/superfluid state of high-$T_c$ cuprates and
related systems and then the vortex-like Nernst effect and
diamagnetism observed above $T_c$ in high-$T_c$ cuprates and the
vortex-like state observed above the $\lambda$-transition
temperature $T_{\lambda}$ in liquid $^4$He.

So far, most researchers confuse the $s$-, $p$- and $d$-wave
pairing states of fermionic quasiparticles in high-$T_c$ cuprates
and other exotic systems with the unusual
superconducting/superfluid states. Therefore, in many cases the
origins of the gapless superconductivity and superfluidity and
gapless excitations, which are manifested in the power law
temperature dependences of the superconducting/superfluid
properties of high-$T_c$ cuprates and other systems, are rashly
attributed to the nodes of $p$- and $d$-wave BCS-like gaps. From
the above considerations, it follows that the unconventional
superconductor/superfluid exhibiting a pseudogap behavior above
$T_c$ at $\varepsilon_F<<1$ eV is not in the fermionic limit of
Cooper pairs but in the bosonic limit of Cooper pairs. We argue
that the gapless superconductivity/superfluidity in high-$T_c$
cuprates and other related systems is associated with the gapless
excitation spectrum of a superfluid 3D Bose-liquid of Cooper pairs
below $T_c^*$ and could not be explicable by the presence of the
point or line nodes of the BCS-like gap. Actually, the BCS-like
$d$-wave pairing model fails to explain the $\lambda$-like
superconducting transition at $T_c$ and the experimental
observation of the existence of gapless excitations below some
characteristic temperature $T_c^*<<T_c$ \cite{67} and their
nonexistence above $T_c^*$ up to $T_c$ in high-$T_c$ cuprates.

We now return to the issue of the vortex-like excitations above
$T_c$. In 3D high-$T_c$ cuprates with $m_p\simeq2.1m_e$,
$m_B=2m_p$, $m^*_B\simeq1.05m_B$ and
$\rho_B\simeq4.2\times10^{19}cm^{-3}$, we find $T_{BEC}\simeq80$
K. For quasi-2D grain boundaries in these systems, we use the
values of $m_p\simeq3m_e$, $m_B=2m_p$, $m^*_B\simeq1.1m_B$ and
$\rho_B\simeq2.5\times10^{13}cm^{-2}$. We then obtain
$T_0\simeq210$ K. Further, we find $T^{3D}_c\simeq
1.135T_{BEC}\simeq91$ K for $\gamma_B=0.3$ and
$\varepsilon_{BA}/k_BT_{BEC}=10$. By taking $\gamma_B=0.3$ for
quasi-2D grain boundaries, we find
$T^{2D}_c\simeq0.68T_0\simeq143$ K. Within the superfluid
Bose-liquid model in the mean-field approximation, thus the
highest $T_c$ is expected to arise in quasi-2D Bose systems. It
follows that the superconducting transition temperature in the
cuprates is higher at quasi-2D grain boundaries than in the bulk
and the residual superconductivity persists at quasi-2D grain
boundaries in the temperature range $T_c<T<T_v(=T^{2D}_c)$, i.e.,
the stability of high-$T_c$ superconductivity in cuprates is
greater in quasi-2D than in 3D systems. Therefore, the vortex-like
Nernst signals observed in high-$T_c$ cuprates \cite{10,68,69} are
caused by the destruction of the bulk superconductivity in the
3D-to-2D crossover region and are associated with the existence of
superconductivity at quasi-2D grain boundaries rather than with
other effects. There is some confusion in the literature
concerning the origins of the vortex-like and diamagnetic states,
which have been found in unconventional cuprate superconductors
above $T_c$ \cite{19,20,68,69}. We argue that the vortex-like
Nernst signals is not associated with the diamagnetic signal
persisting above $T_c$, since the vortex-like state should persist
up to superconducting transition temperature $T^{2D}_c=T_v$ at
quasi-2D grain boundaries, while the diamagnetism above $T_c$ is
associated with the formation of bosonic Cooper pairs (with zero
spin) and would persist up to pseudogap temperature $T^*\gtrsim
T_v$. Another grain boundary effect is that the gap $\Delta_g$ in
the excitation spectrum of a 2D superfluid Bose condensate at
$T\neq0$ is larger than that in the excitation spectrum of a 3D
superfluid Bose condensate at $T>T^*_c$. Hence, the half-integer
$h/4e$ magnetic flux quantization is better manifested in the
3D-to-2D crossover region than in the bulk, as indeed observed at
quasi-2D grain boundaries and in thin films of high-$T_c$ cuprates
\cite{6}, where the half-quantum vortices are associated with the
excitations of pair condensate of bosonic Cooper pairs. Similarly,
the new vortex topology in thin $^4$He superfluid film on porous
media might be intermediate between the bulk superfluid liquid and
flat superfluid film configuration, as discussed in Ref.
\cite{70}. This vortex-like state existing at temperatures
$T_{\lambda}<T<T_c^{2D}$ can be also interpreted as a result of
the crossover from 3D to 2D nature of the superfluid state and
formation of 3D vortices at the destruction of the bulk
superfluidity in the 3D-to-2D crossover region (i.e., in thin
$^4$He film on porous substrate).

\section{The full phase diagram of the normal and superconducting states of high-$T_c$ cuprates}

The undoped cuprate compounds are antiferromagnetic (AF)
insulators. Because the strong electron correlations (i.e., the
strong Coulomb interactions between two holes on the same copper
sites) drive these systems into the AF Mott insulating state.
However, the strong Coulomb interactions of the lattice scale
disappear in doped cuprates \cite{13}. The distinctive feature of
the doped cuprates is the polarizability of their crystal lattice
in the presence of charge carriers introduced by doping. The
self-trapping and pairing of doping carriers are more favorable in
such polar materials than in non-polar solids. In the lightly
doped cuprates, the strong and unconventional electron-phonon
interactions are responsible for the existence of localized
carriers and (bi)polaronic insulating state. Actually, a small
level of doping (e.g., $x=n/n_a\simeq0.02-0.03$ \cite{33,61},
where $n$ is the density of doping carriers, $n_a=1/V_a$ is the
density of the lattice atoms, $V_a$ is the volume per $\rm{CuO_2}$
unit in the cuprates) results in the disappearance of AF order,
the system undergoes a transition from the AF insulator to the
(bi)polaronic insulator. Upon further doping, the cuprate
compounds are converted into a pseudogap metal (above $T_c$) or a
non-BCS high-$T_c$ superconductor (below $T_c$).

\begin{figure}[!htp]
\begin{center}
\includegraphics[width=0.45\textwidth]{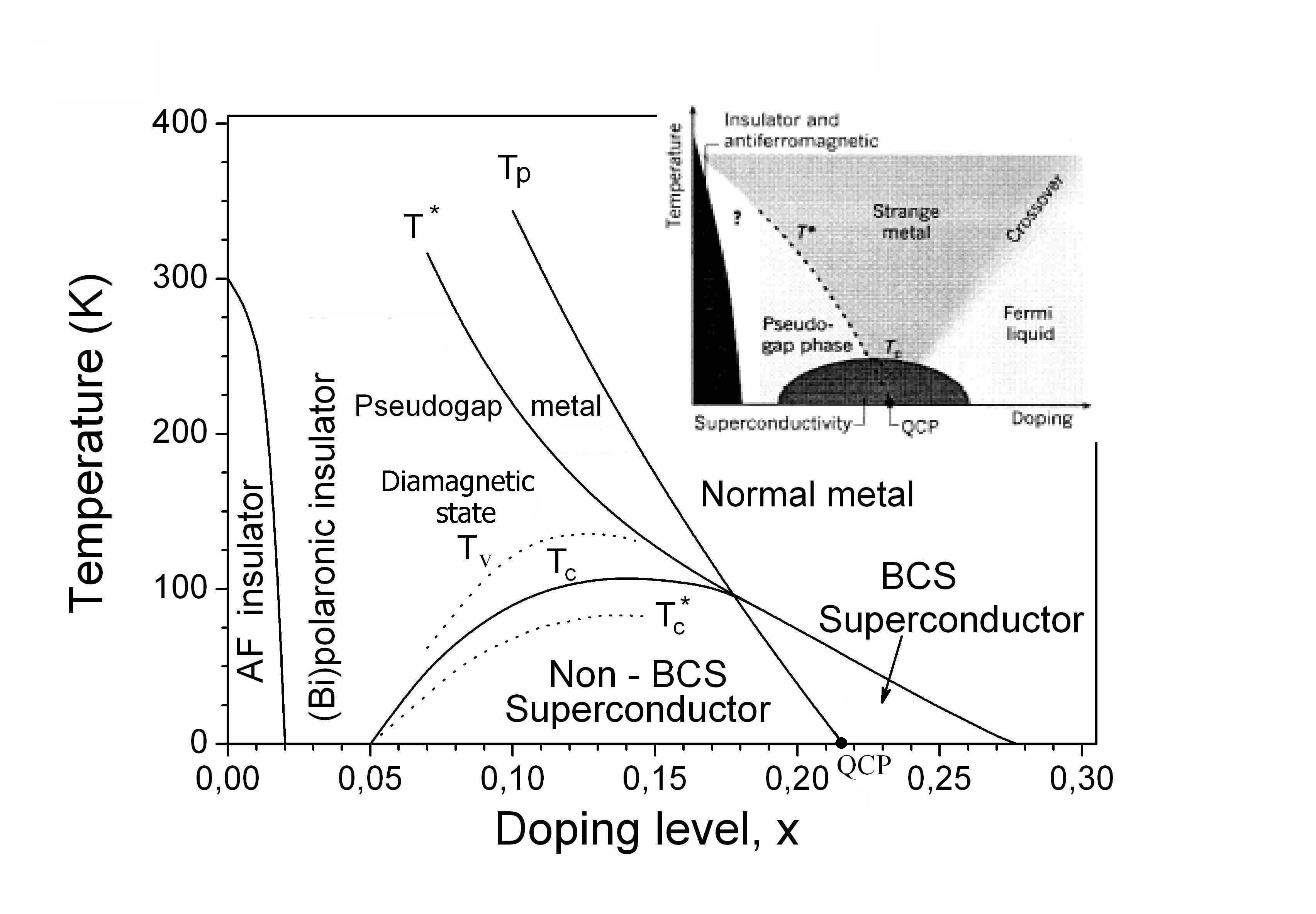}
\caption{\label{fig.5} The full phase diagram of the normal and
superconducting states of Bi-2212 showing various characteristic
temperatures, $T_p$ (the pseudogap phase boundary ending at the
quantum critical point (QCP) \cite{37}), $T^*$ (the BCS-like
pseudogap formation temperature \cite{37}), $T_v$ (the onset of
vortex-like excitations above the bulk superconducting transition
temperature $T_c$) and $T^*_c$ (the onset of the first-order phase
transition in the superconducting state), is compared with the
other phase diagram \cite{11} (see inset). The onset temperature
of vortex formation $T_v$ is higher than $T_c$ but lower than the
onset temperature $T^*$ of diamagnetism in the pseudogap state.
The cuprate superconductor Bi-2212 undergoes a transition from the
Fermi-liquid (BCS-type) superconducting state to the Bose-liquid
one at the QCP ($x\simeq0.22$) and this transition will be
manifested as the normal metal-pseudogap metal transition when
$H=H_{c2}$.}
\end{center}
\end{figure}

The above results show that the high-$T_c$ cuprates are
characterized by low density of condensing (attracting) bosons
$\rho_B<<n$. Here the true superconducting transition temperature
$T_c$ (the onset temperature of the $\lambda$-like second order
phase transition) is determined by postulating that
superconductivity in these systems originates from the superfluid
condensation of a fraction of the normal-state Cooper pairs and is
associated with a microscopic separation between superfluid and
normal bosonic carriers. Such a microscopic phase separation will
likely occur just like the phase separation into the regions of a
Bose solid (high-density limit) and a dilute Bose gas (low-density
limit) described in Ref. \cite{71}. The values of $T_c$ in non-BCS
cuprate superconductors are actually determined by low densities
of bosons and only a part of preformed Cooper pairs is involved in
the superfluid Bose condensation. In 3D systems, the density of
condensing (attracting) bosons is related to $n$ as
$\rho_B=f_sn<<n$, where $f_s$ is the fraction of superfluid
bosons. According to Eqs. (\ref{Eq.14}) and (\ref{Eq.15}), $T_c$
first increases nearly as $T_c\sim(f_sn_ax)^{2/3}$ (in the 3D
case) and $T_c\sim(f^{2D}_{s}n_ax)$ (in the 2D case), then reaches
the maximum at optimal doping and exhibits the saturating or
decreasing tendency with increase of $x$ and $m^*_B$. Thus, both
curves $T^{3D}_c(x)$ and $T^{2D}_c(x)$ have a dome-like shape. A
general advantage of quasi-2D versus 3D systems predicted by the
superfluid Bose-liquid model is that superconductivity can be
observed in a wider region of the phase diagram in the former than
in the latter. The normal state of high-$T_c$ cuprates exhibits a
pseudogap behavior \cite{37}. Further, the onset temperature of
the first-order phase transition $T^*_c$ separates two distinct
superconducting phases of 3D high-$T_c$ cuprates, which arise at
pair and single particle condensations of attracting bosonic
Cooper pairs. The entire phase diagram of $\rm{Bi_2Sr_2CaCu_2O_8}$
(Bi-2212) from Mott insulator to the heavily overdoped regime is
shown in Fig. \ref{fig.5}, where the characteristic temperatures
$T^*_c$, $T_c$ and $T_v$ describe three distinct superconducting
regimes, whereas two unusual metallic states exist below the
crossover temperatures $T_p$ and $T^*$. The vortex-like state
exists in the temperature range $T_c<T<T_v$, while the diamagnetic
state persists up to the BCS-like pseudogap formation temperature
$T^*$.

\section{Discussion and conclusions}

In this paper, we have studied the bosonization of Cooper pairs
and novel superconductivity/superfluidity in high-$T_c$ cuprates
and other pseudogap matters, which are characterized by low Fermi
energies $\varepsilon_F<<1eV$. We have shown that these phenomena
would occur in doped cuprate compounds, heavy-fermion and organic
systems, liquid $^3$He and atomic Fermi gases under certain
conditions. The universal and correct criterion for bosonization
of Cooper pairs in such systems is formulated by using the
uncertainty principle. We found that the condition
$\varepsilon_F\lesssim2\varepsilon_A$ is favorable for the
formation of bosonic Cooper pairs and novel
superconducting/superfluid states in low Fermi energy systems. By
the use of this criterion, we might be able to realize an
unconventional superconductor/superfluid not in the BCS-type but
in the type of bosonic Cooper pairs. In this case the
superconductivity/superfluidity is not simply caused by Cooper
pairing on a Fermi surface and the formation of a BCS-like gap
$\Delta_F$ does not necessarily lead the system to a superfluid
state. Because the underlying mechanism of
superconductivity/superfluidity in a Fermi system depends on the
fermionic or bosonic nature of Cooper pairs. When the Fermi energy
$\varepsilon_F$ becomes comparable with the characteristic energy
$\varepsilon_A$ of the effective attraction between fermions,
superconducting and superfluid matters are in the bosonic regime
$\varepsilon_F\lesssim2\varepsilon_A$.

The BCS-type superconductivity/superfluidity would occur in the
fermionic limit of Cooper pairs (see also Ref. \cite{39}) and can
exist in conventional Fermi systems, in particular, in ordinary
metals ($\varepsilon_F/\hbar\omega_D\sim10^2$) and heavily
overdoped cuprates (at $E_p=0$ and
$\varepsilon_F>>\varepsilon_A=\hbar\omega_0$), where, unlike in
unconventional cuprate superconductors, the superconducting state
is characterized by the BCS-like order parameter and the onset
temperature of Cooper pairing $T^*$ coincides with $T_c$. However,
high-$T_c$ cuprates and other unconventional
superconductors/superfluids could be in the bosonic limit of
Cooper pairs and their Fermi energy is so small \cite{36} that the
size of the Cooper pair $a_c$ becomes small and less than the
spatial separation between two Cooper pairs. Hence, the
BCS-Eliashberg-like theory of superfluid Fermi-liquid cannot be
used to elucidate the mechanisms of superconductivity in many
systems at $\varepsilon_F<2\varepsilon_A$. According to the
two-stage Fermi-Bose-liquid model \cite{31,57}, the BCS-like
pairing theory of fermions, applied to high-$T_c$ cuprates and
other related systems, can describe the formation of Cooper pairs
and pseudogap at $T^*>T_c$, but it fails to account for their
novel superconducting/superfluid states and properties. Here we
have demonstrated that novel types of
superconductivity/superfluidity occurring in the bosonic limit of
Cooper pairs exist in high-$T_c$ cuprates and other low Fermi
energy systems. In high-$T_c$ cuprates the bosonic Cooper pairs
(with zero spin) and diamagnetic state are already formed at a
temperature $T^*$ well above $T_c$, but high-$T_c$
superconductivity is only established when the part of such
composite bosons condenses into a Bose superfluid at $T_c$. In
these and other related systems, the formation of a BCS-like
pairing state of fermions is a necessary, but not a sufficient,
condition for the appearance of superconductivity/superfluidity.
Actually, the discussed BCS-like pairing of fermions at $T^*>T_c$
may be considered as a first step toward a more complete treatment
of Bose-type superconductivity/superfluidity in such systems.
Therefore, two main criteria for the occurrence of unconventional
superconductivity/superfluidity in Fermi systems described by a
two-stage Fermi-Bose-liquid model are following: (i) the BCS-like
order parameter $\Delta_F$ should appear first at $T^*>T_c$, and
(ii) the BCS-like order parameter (or energy gap) $\Delta_{F}$ and
the new coherence parameter $\Delta_B$ (defining the boson
superfluid stiffness) should coexist below $T_c$. The latter
criterion is a necessary and sufficient condition of the novel
superconductivity/superfluidity. The above results clearly
demonstrate that superconductivity/superfluidity of bosonic Cooper
pairs just like superfluidity of $^4$He atoms is well described by
the mean field theory of attracting bosons and the true
superconducting/superfluid phase is identified with the coherence
parameter $\Delta_B$ appearing below $T_c$. As the temperature is
decreased, the pair condensation of attracting bosons occurs first
at $T_c$. Further decrease of the temperature leads to their
single particle condensation at $T_c^*$ somewhat below $T_c$ (in
three dimensions) or at $T=0$ (in two dimensions). The gapless
superconductivity/superfluidity occurs below $T_c^*$ due to the
vanishing of the gap $\Delta_g$ in $E_B(k)$ at $T<T_c^*$. The
coherent single particle and pair condensates of bosonic Cooper
pairs exist as the two different superfluid phases and arise from
an effective attraction between these composite bosons in some
domains of momentum space. According to the superfluid Bose-liquid
model, the cuprate high-$T_c$ superconductivity is more robust in
quasi-two-dimensions than in three dimensions, i.e., $T_c$ is
higher in quasi-2D than in 3D systems. We see therefore that,
three different superconducting phases exist in high-$T_c$
cuprates where the coherent pair condensate of bosonic Cooper
pairs persists up to the temperature $T_v=T^{2D}_c>T^{3D}_c$ at
quasi-2D grain boundaries as the superfluid phase and the coherent
pair and single particle condensates of such composite bosons in
3D systems exist as the two distinct superfluid phases below
$T_c=T_c^{3D}$. It follows that the persistence of the vortex-like
excitations in high-$T_c$ cuprates above $T_c$ is caused by the
destruction of the bulk superconductivity. The existence of such
vortices is expected below the temperature $T_v$ lower than $T^*$
but higher than $T_c$. One of the important conclusions is that
diamagnetism in the pseudogap state and vortex formation above
$T_c$ in high-$T_c$ cuprates are unrelated phenomena.

Clearly, the condensate and excitations of a Bose-liquid are
unlike those of a BCS-like Fermi liquid. Therefore, not all the
experimental methods are able to identify the true superconducting
order parameter in high-$T_c$ cuprates and other pseudogap
matters. For example, the single-particle tunneling spectroscopy
and ARPES provide information about the excitations gaps at the
Fermi surface but fail to identify the true superconducting order
parameter appearing below $T_c$ in non-BCS superconductors.
Actually, a prolonged discussion of the origin of unconventional
superconductivity in the cuprates on the basis of tunneling and
ARPES data has nothing to do with the true mechanism of high-$T_c$
cuprate superconductivity. We note here that the
superconducting/superfluid order parameter in high-$T_c$ cuprates
and other pseudogap matters should not be identified as a BCS-like
gap and the gapless superconductivity/superfluidity in these
systems should also not be attributed to the point and line nodes
of the BCS-like ($p$- and $d$-wave) gaps.

There is now experimental evidence that the BCS-like fermionic
excitation gap $\Delta_F$ exists as a pseudogap in high-$T_c$
cuprates \cite{10,12} and other related superconductors
\cite{72,73,74} and atomic Fermi gases \cite{32,75,76}. In non-BCS
superconductors, the superconducting/superfluid order parameter
$\Delta_B$ appearing below $T_c$ and the BCS-like gap $\Delta_F$
opening on the Fermi surfaces above $T_c$ have different origins.
Unconventional high-$T_c$ superconductivity in cuprates is
controlled by the coherence parameter (superfluid stiffness)
$\Delta_B\sim \rho_B$ rather than BCS-like pairing gap
$\Delta_F>>\Delta_B$ and appears under the coexistence of two
order parameters $\Delta_F$ and $\Delta_{SC}(=\Delta_B)$. The
BCS-like pseudogap is therefore a necessary ingredient for
high-$T_c$ superconductivity in the cuprates. Some selected
experimental techniques can provide information about the new
superconducting order parameter. The above presented results show
that the thermodynamic methods and the methods of critical current
and magnetic field measurements are sensitive to the
identification of $\Delta_{SC}(T)$ in unconventional
superconductors.

Thus, the criterion for bosonization of Cooper pairs
$\varepsilon_F\lesssim2\varepsilon_A$ allows us to find the real
applicability boundary (which up to now remains unknown) between
BCS-type and Bose-type regimes of superconductivity/superfluidity.
This criterion and other necessary and sufficient criteria
formulated here should be satisfied for the occurrence of the
unconventional superconductivity/superfluidity. The above
theoretical predictions and their experimental confirmations speak
strongly in favor of the existence of novel
superconducting/superfluid states, which arise in high-$T_c$
cuprates and other pseudogap matters at single particle and pair
condensations of attracting bosonic Cooper pairs. The critical
behavior of a superfluid Bose liquid of Cooper pairs near $T_c$ is
similar to that of liquid $^4$He near the $\lambda$-transition.
Within the mean field theory of a superfluid Bose-liquid, it is
possible to describe the following unexplained features of
unconventional superconductors and superfluids: (i) the key
features of the phase diagrams of high-$T_c$ cuprates (e.g.,
vortex-like state existing at temperatures $T_c<T<T_c^{2D}$ and
two distinct superconducting phases below $T_c$), (ii) the two
distinct superconducting phases in heavy-fermion systems below
$T_c$, (iii) the superfluid $A$ and $B$ phases in $^3$He, (iv) the
superfluid phase in $^4$He below $T_{\lambda}$ and the vortex-like
state existing at temperatures $T_{\lambda}<T<T_c^{2D}$ in the
crossover regime between the bulk superfluid liquid and thin
$^4$He superfluid film, (v) the unconventional superfluidity in
ultracold atomic Fermi gases, (vi) the $\alpha$- clustering in
nuclei (i.e. $\alpha$-particle structure of nuclei) and high
stability of magic and twice magic nuclei, which is associated
with the single particle and pair condensations of attracting
bosonic Cooper pairs of nucleons both in proton and in neutron
subsystems. We have shown that this theory provides a consistent
picture of the distinctive superconducting properties (e.g.,
$\lambda$-like second-order phase transition at $T_c$, first-order
phase transition and kink-like temperature dependences of
superconducting parameters near $T_c^*$) of high-$T_c$ cuprates
and other related materials.

\section*{Acknowledgments}

I gratefully acknowledge discussions with J. Zaanen, H. Koizumi,
A.L. Solovjov, P.J. Baimatov, E.M. Ibragimova, B.L. Oksengendler,
U.T. Kurbanov, Z.S. Khudayberdiev and E.X. Karimbaev. This work
was supported by the Foundation of the Fundamental Research, Grant
No $\Phi$2-$\Phi$A-$\Phi$120.

\end{document}